\documentclass[aps,pre,twocolumn,groupedaddress,floatfix,showpacs]{revtex4}

\usepackage{graphicx}
\usepackage{latexsym}

\begin{document}

\begin{widetext}
\noindent\textbf{Preprint of:}\\
Wolfgang Singer, Timo A. Nieminen, Ursula J. Gibson,
Norman R. Heckenberg and Halina Rubinsztein-Dunlop\\
``Orientation of optically trapped nonspherical birefringent particles''\\
\textit{Physical Review E} \textbf{73}(2), 021911 (2006)
\end{widetext}

\title{Orientation of optically trapped nonspherical birefringent particles} 

\author{Wolfgang Singer}
\email[]{singer@physics.uq.edu.au}
\author{Timo A. Nieminen}
\email[]{timo@physics.uq.edu.au}
\author{Ursula J. Gibson}
\altaffiliation{Permanent address:
Thayer School of Engineering, Dartmouth College,
Hanover NH 03755-8000}
\author{Norman R. Heckenberg}
\author{Halina Rubinsztein-Dunlop}

\affiliation{Centre for Biophotonics and Laser Science, Department of Physics,
The University of Queensland, Brisbane QLD 4072, Australia}

\begin{abstract}
While the alignment and rotation of microparticles in optical traps
has received increased attention recently, one of the earliest
examples has been almost totally neglected---the
alignment of particles relative to the beam axis, as opposed to about
the beam axis. However, since the alignment torques determine how particles
align in a trap, they are directly relevant to practical applications.
Lysozyme crystals are an ideal model system to study factors determining
the orientation of nonspherical birefringent particles in a trap.
Both their size and their aspect ratio can be controlled by the growth
parameters, and their regular shape makes computational modelling
feasible. We show that both external (shape) and internal (birefringence)
anisotropy contribute to the alignment torque. Three-dimensionally
trapped elongated objects either align with their long axis parallel
or perpendicular to the beam axis depending on their size. The
shape-dependent torque can exceed the torque due to birefringence,
and can align negative uniaxial particles with their optic axis
parallel to the electric field, allowing application of optical torque
about the beam axis.
\end{abstract}

\maketitle 

\section{Introduction}

Optical tweezers have been used to manipulate and
investigate microscopic particles for many years,
and a wide variety of applications have been
explored. The underlying principle behind optical
tweezers is the transfer of momentum from the trapping
beam to the particle~\cite{ashkin1986}. As light can carry
angular momentum as well as linear momentum, 
torque can also be exerted on particles
in optical tweezers.

The optical torque acting about the beam axis is always a
result of the alteration of orbital and/or spin angular
momentum of the incident beam by the trapped particle,
by absorption or by scattering if there is either
external (shape) or internal
(birefringence) anisotropy~\cite{nieminen2004c}.
Consequently, the torque can
either originate from a beam where the incident light itself
carries angular momentum that is transferred to the particle,
or it can originate from a beam where the incident light
carries zero angular momentum, but where the trapped particle
induces angular momentum in the beam.

A variety of methods
to accomplish angular momentum transfer have been proposed
and tested~\cite{friese1996pra,galajda2001,%
bishop2003}.
Of these, one of the best
suited for actual practical applications is the transfer
of incident spin angular momentum to birefringent
particles~\cite{friese1998nature,bishop2004,laporta2004}.
Firstly, spin angular momentum can easily be measured,
so that the applied optical torque can be determined by
purely optical means, making the system well suited for
quantitative measurements~\cite{nieminen2001jmo,bishop2004}.
Secondly, the torque
can be controlled by changing the polarization state of
the light, keeping the power constant.
Thirdly, the torque is quite high, typically on the order
of $\hbar$ per photon per second if highly birefringent particles
are used. Finally, this method can be used with Gaussian
beams, ensuring high 3D trapping efficiency. 

However, in order to act as a wave plate, the birefringent
particle cannot be oriented with the optic axis parallel
to the beam axis. Only for other orientations
is the polarization state
of the light altered and spin angular momentum
transferred from the beam to the particle, causing either
constant particle rotation in circularly polarized light or
particle alignment in linearly polarized light.
Obviously, maintaining the required orientation
is crucial for the use of birefringent particles
as micromotors or for other applications requiring rotation.

Similar principles apply to flattened or elongated particles
which also can alter the angular momentum of the incident
light only if their asymmetry about the beam axis is conserved after
being trapped.
Orientation effects due to the shape of the particle have been
reported previously. In particular, it has long been known that
elongated particles tend to align with their long axis
along the axis of the trapping beam~\cite{ashkin1987b}.
However, this behaviour is not universal, and the orientation
of trapped particles depends on their size, shape, and optical
properties~\cite{nieminen2001f}.

The torques giving rise to the orientation of particles with
respect to the beam axis have received little attention; this is
due in part to the transient nature of the torques, which act to
align the particle when it is trapped, and also to the difficulty
of calculating torques on nonspherical particles.
Earlier work has usually made use of the geometric optics
approximation~\cite{gauthier1997b}, or been restricted to
particles of simple geometry and homogeneous and isotropic
material~\cite{bayoudh2003}. Extension of these foundations
to smaller particles for which geometric optics fails, and
to more complex particles is highly desirable.

Here we will
present numerical calculations---substantiated by experimental
results---on how birefringent
particles with different aspect ratios and
radii will align after being three-dimensionally trapped.
Moreover, we are completing the picture of shape-dependent
alignment of particles, showing that elongated particles can
align with their long axis either parallel or perpendicular
to the beam axis, depending on their aspect ratio and
size compared to the beam waist. 

However, the internal anisotropy (birefringence) also contributes
to the torque that determines orientation relative to the beam axis.
This torque---like the one due to
the shape---only occurs at the beginning of a trapping event
and has also not yet been studied in detail.  Our results
reveal that the birefringence-induced torque responsible for alignment
to the beam axis has the
same order of magnitude as the birefringence-induced torque
about the beam axis. This is in contrast to the shape-induced
torque which can be an order of magnitude larger relative to the
beam axis as compared to the one about the
beam axis~\cite{bayoudh2003}.

We show both experimentally and by computational modelling
that, depending on the respective anisotropy, one aligning
effect can dominate the other. Using growing lysozyme
crystals we were able to observe the transition where
the alignment torque due to birefringence overcomes the torque
due to shape when we changed the aspect ratio of a particle
while trapped.  These findings explain how
negative uniaxial calcite crystals
can be spun in circularly polarized light~\cite{friese1998nature},
despite the fact that such crystals tend to align with their
optic axis parallel to the beam axis~\cite{laporta2004}.

\section{Orientation of lysozyme crystals}

Lysozyme crystals are a widely used model to study nucleation
and growth of protein crystals. In its most common
(tetragonal) form, lysozyme forms a positive uniaxial
birefringent crystal, with a well-characterized morphology.
A schematic drawing showing the crystallographic axes and
faces can be found in reference~\cite{singer2005}. The optic axis,
which coincides with the crystal's c-axis, points from one tip
to the other tip of the crystal. The difference in the indices
along the [001] and [110] axes at 1064\,nm is
$1.66\times 10^{-3}$~\cite{singer2004}. The relative
ordinary refractive
index ($m = n_\mathrm{crystal}/n_\mathrm{medium}$) was found to be
$1.039$ from the measured trap stiffness~\cite{knoner}. The aspect
ratio is defined as the length along the optic axis to the width of
the [110] face.

The size and shape distributions of lysozyme crystals can be
controlled by varying the initial salt and protein concentrations.
This allows
the experimental study of the shape-dependence of the
orientation of optically trapped particles. The size and
aspect ratio can even be changed dynamically, while the lysozyme
crystal is held in the optical trap.
 
The experiments were carried out using a setup described in detail
by Singer et al.~\cite{singer2004}. Briefly, linearly polarized
light from an Yb-doped fiber laser operating at 1070\,nm in the
range of 200--500\,mW was coupled into a $60\times$ oil-immersion objective
of numerical aperture 1.4, with a resulting beam waist radius of
$0.44\,\mu$m. A half-wave plate in the beam path could
be used to adjust the direction of the plane of polarization of the
trapping beam.
The exact alignment of the individual particles in the trap could
be determined using a software package developed by
Gibson and Kou~\cite{gibson}.


We calculated the optical force and torque acting on lysozyme
crystals in an optical trap by using the \textit{T}-matrix
method to determine the scattering of the trapping beam by
the crystal, and finding the difference between the inflow and
outflow of electromagnetic momentum and angular momentum,
which is equal to the force and torque exerted on the
crystal~\cite{nieminen2004d,nieminen2003a,nieminen2003b,bishop2003}.

Since calculation of the \textit{T}-matrix is much faster for
axisymmetric particles~\cite{waterman1971}, the lysozyme crystals
were modelled as cone-tipped cylinders. Since their relative
refractive index is close to 1, and their birefringence is small,
even compared with this small refractive index contrast, it was
possible to assume an effective isotropic refractive index
(which is a function of the orientation of the crystal) to
separately calculate the torque due to the shape. The torque due
to the birefringence was then calculated using the angle-dependence
of the orientation energy of the birefringent material in the applied
field; this procedure, while approximate, is valid in the low-contrast
low-birefringence case of lysozyme crystals.

\begin{figure*}[tb]
\includegraphics[width=0.9\textwidth]{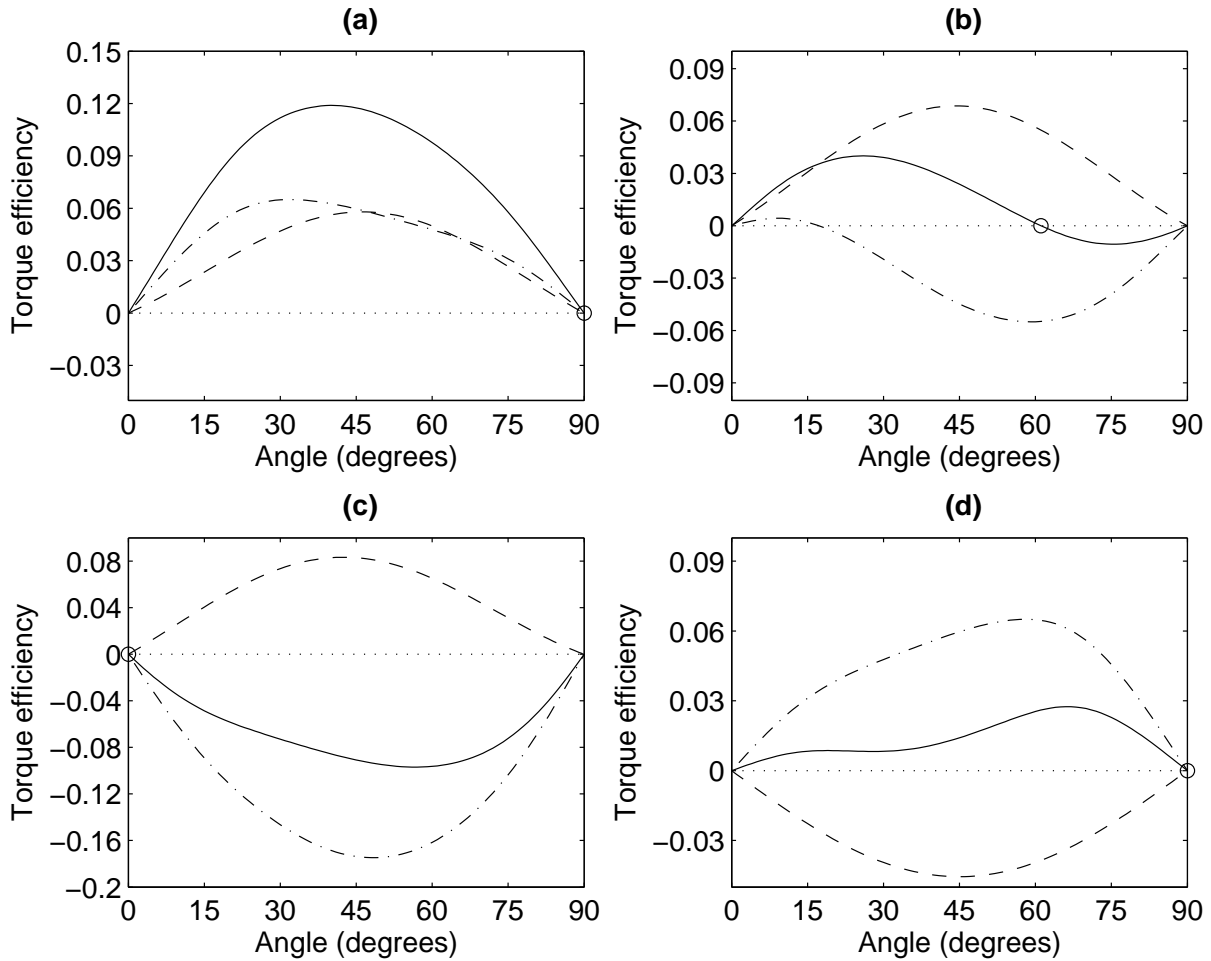}
\caption{Torque (shown as torque efficiency, in $\hbar$/photon)
acting on trapped lysozyme crystals.
Solid lines show the total
torque, dot--dash lines the torque due to shape alone, and dashed
lines the torque due to birefringence alone. The dotted line indicates
zero torque; equilibrium points occur when the total torque curve
(solid line) crossed this dotted line, with the stable equilibria
indicated by the small circles. In all cases, since the crystals
are positive uniaxial, the torque acts to align the optic axis
(the symmetry axis of the model crystals) with the electric field.
(a)--(c)
show the dependence of the torque on the angle between the
optic axis of the crystal and the beam axis. A positive torque acts
to align the optic axis with the beam axis.
The torque is shown for crystals
of radius 3\,$\mu$m and aspect ratios of (a) 0.9, (b) 1.2, and
(c) 1.6. Three different equilibrium orientations can be seen:
optic axis perpendicular to the beam axis (a), with both the torque
due to birefringence and the torque due to shape acting in the same
direction; parallel to the beam axis (c), with the two contributions
to the torque opposing each other, but with the shape contribution
dominant; and at an intermediate angle (b). In (b), if the shape torque
was the only torque acting, the crystal would still align at an
intermediate angle. The birefringence torque still acts to align
the optic axis perpendicular to the beam axis and shifts the equilibrium
angle closer to being perpendicular to the beam axis, but is insufficient
to overcome the shape torque.
The torque
about the beam axis acting on the crystal in (a) is shown in (d).
A positive torque acts to align the optic axis with the plane of
polarisation; the angle is the angle between the optic axis and the
plane of polarisation.
The shape torque and birefringence torques oppose each other. In this
case, the shape torque is dominant; for larger crystals, the birefringence
torque will increase approximately proportional to the radius, while the
shape torque will decrease~\cite{bishop2003}}
\label{torque_angle}
\end{figure*}

Equilibrium orientations were found from the dependence of torque
on the angle between the axis of the crystal model and the beam axis.
The torque versus angle for different aspect ratios
is shown in figure~\ref{torque_angle}. Three
distinct regimes of behaviour---alignment with the crystal optic axis
perpendicular to the beam axis, parallel to the beam axis, and
at an intermediate angle---can be identified.
This behaviour can be understood in terms of the usual behaviour of
nonspherical particles in optical traps---such particles tend to align
with their longest dimension along the beam axis. Therefore, a high-aspect
ratio lysozyme crystal (elongated along the optic axis)
will align with the optic axis parallel to the beam axis. For
crystals with smaller aspect ratios the body diagonal is the longest
axis and alignment with the body diagonal parallel to the beam axis
is to be expected, leading to the optic axis being skewed to the
beam axis by an angle depending on the aspect ratio~\cite{singer2005}.
This type of alignment is widely observed with flattened
particles~\cite{bayoudh2003}.
Meanwhile, since the crystals are positive uniaxial, the torque due
to birefringence acts to align the optic axis perpendicular to the
beam axis. For low aspect ratio (flattened) particles, both the
torque due to shape and the torque due to birefringence act in the
same direction, and the crystal aligns with the optic axis perpendicular
to the beam axis, as seen in figure 1(a). For elongated crystals,
these torques oppose each other, and the transition to alignment along
the beam axis requires a larger aspect ratio than would be the case
without birefringence.

However, this simple picture fails to explain the existence of the third
regime---alignment with the optic axis perpendicular to the beam
axis. Notably, the shape torque alone would result in alignment at an
intermediate angle, at an angle of $17^\circ$. Since the birefringence
torque acts to align the optic axis perpendicular to the beam axis,
but is insufficient to completely overcome the torque due to shape,
this angle is increased to $61^\circ$.

\begin{figure}[!htbp]
\centerline{\includegraphics[width=0.97\columnwidth]{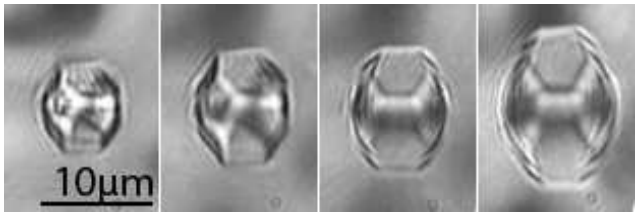}}
\caption{Growth of a trapped lysozyme crystal. The aspect
ratio can be changed during growth, altering the orientation
of the trapped crystal.}
\label{growthfig}
\end{figure}

\begin{figure}[!htbp]
\centerline{\includegraphics[width=\columnwidth]{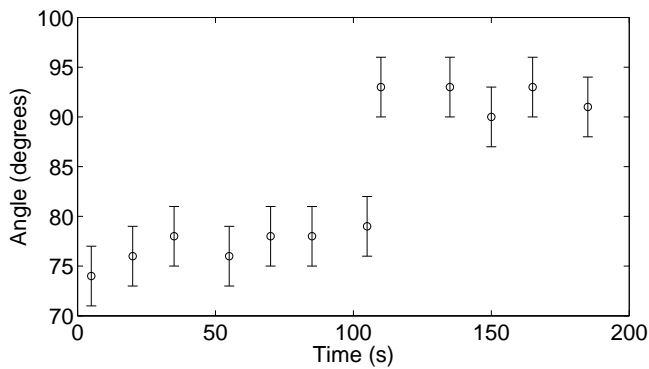}}
\caption{Change in orientation of a growing lysozyme crystal. The
angle between the optic axis of the crystal and the beam axis is shown.}
\label{angle_timefig}
\end{figure}

\begin{figure}[!htbp]
\centerline{\includegraphics[width=\columnwidth]{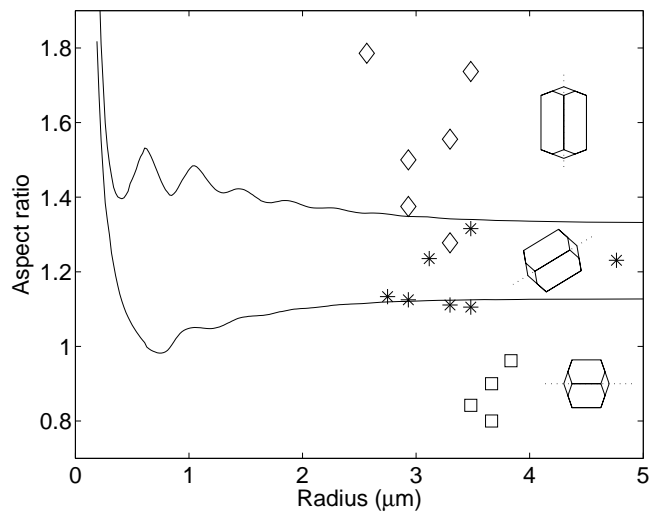}}
\caption{Equilibrium orientation of trapped lysozyme crystals with
different aspect ratios and sizes. The solid lines represent the
calculated borders between the different regimes; the orientation
within each regime is shown by the inset crystal profiles, showing
a side-view of the crystal, relative to a vertical trapping beam with
plane polarization in the plane of the page. Observed orientations
are indicated by: $\Diamond$---optic axis parallel to the beam axis,
$\Box$---perpendicular, and {\large$\ast$}---intermediate. The
top-view photographs shown in figure 2 correspond to the intermediate
(first two frames) and the perpendicular (last two frames) cases.}
\label{orientationfig}
\end{figure}

We observed all three types of alignment with lysozyme crystals of
varying size and aspect ratio. Since it is possible to grow crystals
while they are trapped, we were able to observe the transitions between
these regimes of alignment in individual crystals. Figure~\ref{growthfig}
shows the growth of a trapped crystal
while the protein concentration in the solution was changed,
with a change in orientation between frames 2 and 3. It
can be seen that for the given growing conditions the crystal is
growing primarily by addition of material on the (110) faces, as
would be expected at a high protein concentration~\cite{forsythe1999}. 
The measured angle between the optic axis of the crystal and
the beam axis is shown in figure~\ref{angle_timefig}.
The transition to the perpendicular alignment occurs when the
torque due to the birefringence of the crystal becomes dominant
over the shape-dependent torque. However, even if the crystals were
optically isotropic, this transition would still occur (though at a
smaller aspect ratio) when the shape-dependent torque changes direction.

The observed equilibrium orientations of a number of lysozyme crystals
of varying size and aspect ratio are shown in figure~\ref{orientationfig}.
The calculated extents of the different regimes of orientation
are also shown, and agree well with the observed orientations.
The crystals were all in the same sample. Note that the crystals drawn
in figure 4 to show the equilibrium orientation in each regime
are shown in side view, while figure 2 shows their appearance
when viewed in the microscope (ie a top view).



\section{Other particles}

While the previous results apply specifically to
lysozyme crystals, which have a low refractive index
contrast with the medium and a small birefringence, the same
general principles apply to other particles as well.
While the shape-dependent torque varies with the refractive
index of the particle (proportional to the refractive index
contrast $m-1$ in the low-contrast limit), the orientations for
which the torque is zero only weakly depends on the refractive
index---a refractive index contrast ten times larger 
yields boundaries between the regimes of orientation very similar
to those in figure~\ref{orientationfig}.

This is especially relevant when we consider the trapping and
rotation about the beam axis of birefringent particles. If the
particle is positive uniaxial, the birefringence-dependent torque
acts to align the optic axis perpendicular to the beam axis. The
torque on a negative uniaxial particle, on the other hand, acts to
align the optic axis with the beam axis. If this is the equilibrium
orientation of the particle, the particle appears to be isotropic
as far as the incident beam is concerned, and no transfer
of angular momentum occurs~\cite{laporta2004,nieminen2004c}.
This raises the question of why optically trapped
negative uniaxial crystals have been observed to
spin~\cite{friese1998nature}.
It appears reasonable to suppose that, especially since the
crystals in question were irregular, shape-dependent
torques produced an equilibrium orientation such that the
optic axis was not parallel to the beam axis.

It can also be seen in figure~\ref{orientationfig} that,
if the particle is small compared to the beam waist, elongated
particles can align with their long axis perpendicular to
the beam axis, in agreement with previous results for very small
particles~\cite{nieminen2001f}.
These findings can be explained
by the fact that particles that are small compared to the beam
waist are trapped in the centre of the focal spot, where the
intensity gradient is small. The elongated objects
therefore align with the axis of the highest particle
polarisability---their longest axis---in the direction of the
electric field vector, and therefore perpendicular to the beam
axis, rather than parallel to the beam axis.

\section{Conclusions}

Of the several methods to orient and rotate microscopic
particles in optical tweezers, by the far the most important
to date, as far as quantitative measurements are concerned,
is transfer of spin angular momentum to birefringent
particles~\cite{nieminen2001jmo,bishop2004,laporta2004}.
However, the ability to use birefringent
particles is restricted to those which orient with their optic axis
not parallel to the beam axis after being trapped, which at
first appears to rule out the use of negative uniaxial materials.
We have shown that the torque due to nonspherical shape can overcome
the torque due to birefringence, and can be used to maintain negative
uniaxial particles in the desired orientation.
Furthermore, we showed that elongated particles small compared to the beam
waist will align perpendicular to the beam axis.

The results
presented are relevant to the design of particles that can
be used as motors in optically-driven micromachines, and have
potential to increase the range of particles that can
serve to probe properties of microscopic or biological systems.
The predictability and computability of these torques enables their
practical use in optical micromanipulation.

\section{Acknowledgements}

We would like to acknowledge the support of NASA grant NAG8-1590,
the University of Queensland and the Australian Research Council.
We are indebted to Gregor Kn\"{o}ner and Simon Parkin for their
contributions.


\end{document}